\documentclass[twocolumn,aps,prl,groupedaddress,footinbib,amsmath,amssymb,showpacs]{revtex4-1}
\usepackage{graphicx}
\usepackage{bm}
\usepackage[colorlinks,	linkcolor=red,anchorcolor=blue,citecolor=blue,bookmarksnumbered]{hyperref}
\usepackage{float}

\begin{document}
	
	\title{Self-induced Transparency in Warm and Strongly Interacting Rydberg Gases}
	\author{Zhengyang Bai$^{1,2}$, Charles S. Adams$^{3}$, Guoxiang Huang$^2$ and Weibin Li$^{1}$}
	\affiliation{
		$^1$School of Physics and Astronomy, and Centre for the Mathematics and Theoretical Physics of Quantum Non-equilibrium Systems, University of Nottingham, Nottingham, NG7 2RD, UK\\
		$^2$State Key Laboratory of Precision Spectroscopy, East China Normal University, Shanghai 200062, China\\
		$^3$Joint Quantum Centre (JQC) Durham–Newcastle, Department of Physics, Durham University, South Road, Durham, DH1 3LE, United Kingdom
	}
\begin{abstract}
	We study dispersive optical nonlinearities of short pulses propagating in high number density, warm atomic vapors where the laser resonantly excites atoms to Rydberg $P$-states via a single-photon transition. Three different regimes of the light-atom interaction, dominated by either Doppler broadening, Rydberg atom interactions, or decay due to thermal collisions between groundstate and Rydberg atoms, are described. We show that using fast Rabi flopping and strong Rydberg atom interactions, both in the order of gigahertz, can overcome the Doppler effect as well as collisional decay, leading to a sizable dispersive optical nonlinearity on nanosecond timescales. In this regime, self-induced transparency (SIT) emerges when areas of the nanosecond pulse are determined primarily by the Rydberg atom interaction, rather than the area theorem of interaction-free SIT. We identify, both numerically and analytically, the condition to realize Rydberg-SIT. Our study
	contributes to efforts in achieving quantum information processing using glass cell technologies.
\end{abstract}

\maketitle

\textit{\textbf{Introduction.}---}
Strong and long-range interactions between atoms excited in high-lying Rydberg states~\cite{jaksch_fast_2000,lukin_dipole_2001,saffman_quantum_2010} can be mapped onto weak light fields via electromagnetically induced transparency (EIT)~\cite{pritchard_cooperative_2010,petrosyan_electromagnetically_2011,li_2014,firstenberg_nonlinear_2016,sevincli_nonlocal_2011,bai_enhanced_2016,bai_stable_2019}, permitting interaction-mediate optical nonlinearities~\cite{gorshkov_photon-photon_2011,peyronel_quantum_2012,gorshkov_dissipative_2013,firstenberg_attractive_2013,he_two-photon_2014,busche_contactless_2017,liang_observation_2018} and optical quantum information processing~\cite{maxwell_storage_2013,li_quantum_2016,tiarks_optical_2016,tiarks_photonphoton_2019,gorniaczyk_single_2014,baur_single-photon_2014,li_coherence_2015,tiarks_single-photon_2014,gorniaczyk_enhancement_2016,murray_coherent_2017}. In the EIT approach, ultracold temperatures ($\sim \mu$K) are of critical importance to maintain the dispersive nonlinearity (typically sub-megahertz). As Doppler broadening ($\propto \sqrt{T}$ with $T$ the temperature) increases from about 100 kilohertz at $1\,\mu$K to gigahertz at $300$ K, large thermal fluctuations at high temperatures can easily smear out the nonlinearity~\cite{carr_nonequilibrium_2013,baluktsian_evidence_2013,urvoy_strongly_2015,zhang_nonlocal_2016}. To overcome this limitation, recent experiments employ short (nanoseconds) and strong (gigahertz Rabi frequencies) lasers to excite high density, room-temperature (or hot) Rydberg gases~\cite{baluktsian_evidence_2013,urvoy_strongly_2015,ripka_room-temperature_2018} confined in glass cells~\cite{briaudeau_coherent_1998,fichet_exploring_2007,sarkisyan_spectroscopy_2004,moiseev_complete_2001}. Through a four-wave mixing process, strong dispersive nonlinearities even exceed the laser strength and thermal effect to realize a single photon source in the glass cell setting~\cite{ripka_room-temperature_2018}. Though rapid experimental developments~\cite{baluktsian_evidence_2013,urvoy_strongly_2015,ripka_room-temperature_2018}, theoretical understanding of the optical nonlinearity mediated by Rydberg interactions that emerges in nanosecond timescale and room temperature gases remains unavailable.
\begin{figure}
	\centering
	\includegraphics[width=0.45\textwidth]{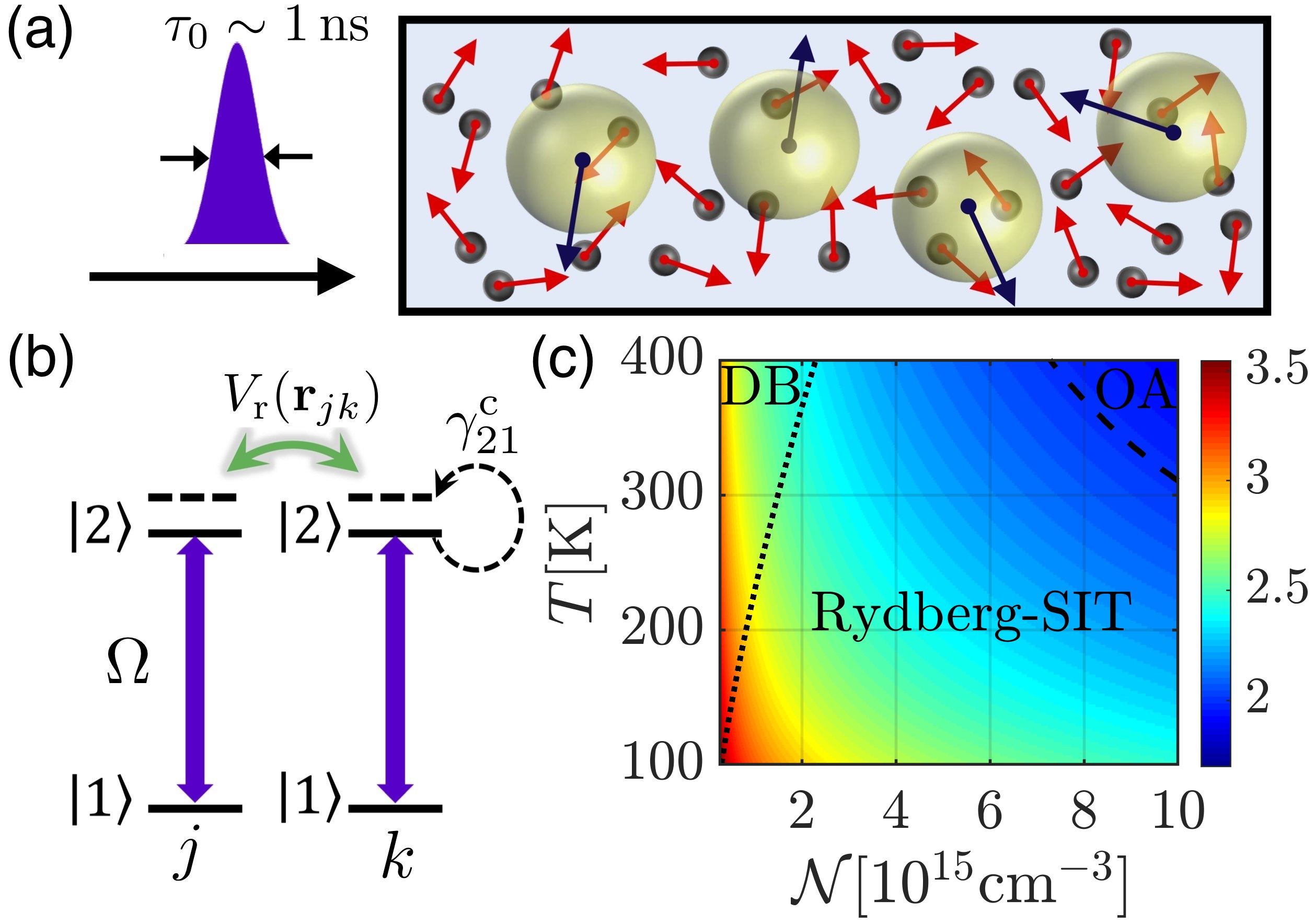}\\
	\caption{\footnotesize(Color online) \textbf{Light-atom interactions in thermal gases.} (a)~Nanosecond pulses excite thermal atoms to Rydberg states. This process is affected by thermal motions, Rydberg atom interactions, and inelastic collisions between groundstate (black dots) and Rydberg (yellow balls) atoms. (b)~Level scheme. The laser (Rabi frequency $\Omega$) resonantly couples groundstate $|1\rangle$ and Rydberg state $|2\rangle$. The latter experiences strong, long-range van der Waals interactions $V_r({\bf r}_{jk})$ and collisional decay (rate $\gamma_{21}^{\rm c}$). (c)~Three regimes of nanosecond pulses, dominated  by either the Doppler broadening (DB), Rydberg interactions, or optical absorption (OA). Rydberg-SIT forms in the Rydberg interaction dominant region. The color bar shows the ratio of the sum of the Rydberg interaction and Doppler broadening over the collisional decay rate. See text for details.
	}\label{fig1}
\end{figure}

In this work we theoretically investigate dispersive optical nonlinearities of nanosecond light pulses generated in thermal gases of Rydberg atoms excited via a single-photon transition. A crucial requirement to generate significant Rydberg interactions at high temperatures is the high number density of the gas, which can cause large inelastic collisions between groundstate atoms and Rydberg electrons. Using nanosecond pulses, we identify a dispersive nonlinear regime where the Rydberg interaction surpasses the thermal and collisional effects. Importantly this Rydberg nonlinearity depends non-perturbatively on the transient dynamics of the atoms. A key finding in this regime is that Rydberg self-induced transparency (SIT) can emerge in low and high temperature gases, where the pulse shapes into a bright soliton. Through numerical and mean-field calculations, we show explicitly that the formation of Rydberg-SIT depends critically on the Rydberg interaction. This is fundamentally different from conventional (i.e. no two-body interactions) SIT which is governed by the area theorem barely due to light intensities~\cite{mccall_self-induced_1967}. Our study opens opportunities to investigate Rydberg nonlinear optics and possibly to implement optical quantum information processing with warm Rydberg gases in nanosecond timescales.

\textit{\textbf{Light-atom interaction.}---}
We consider nanosecond laser pulses (wave vector $\mathbf{k}$ along the $z$ axis) propagating in a high density gas (density $\cal{N}$), as depicted in Fig.~\ref{fig1}(a). The laser resonantly couples groundstate $|1\rangle$ to Rydberg $nP $  state $|2\rangle$ (with $n$ the principal quantum number) via a single-photon transition [see Fig.~\ref{fig1}(b)]. Two Rydberg atoms (located at ${\bf r}_j$ and ${\bf r}_k$) interact via the van der Waals (vdW) interaction $V_r({\bf r}_{jk})=-C_6/|{\bf r}_{jk}|^6$ with ${\bf r}_{jk}={\bf r}_j-{\bf r}_k$ and $C_6\propto n^{11}$ to be the dispersion coefficient. In this setting, Rydberg electrons frequently collide with surrounding groundstate atoms through the polarization interaction. Using the Fermi pseudo-potential and neglecting higher partial waves~\cite{omont_theory_1977}, such interaction is approximated to be $V_p(\mathbf{r}_{jk})\approx 2\pi a_s \delta(\mathbf{r}_{jk})$~\cite{beigman_collision_1995} where $a_s$ is the s-wave scattering length of the electron-atom collision~\footnote{Energy dependence of the electron-atom scattering is neglected as the average velocity of a Rydberg electron $\sim v_0/n$ is far larger than the the typical velocity of atoms at room temperature.}. This yields the $N$-atom Hamiltonian ($\hbar\equiv 1$)
\begin{eqnarray}
\hat{H}&=&
\sum_{j=1}^N  H_j
+\sum_{k\neq j}^N\left[\frac{V_r(\mathbf{r}_{jk})}{2}\hat{\sigma}^j_{22}\hat{\sigma}^k_{22} + V_p(\mathbf{r}_{jk})\hat{\sigma}^j_{22}\hat{\sigma}^k_{11} \right]\nonumber
\end{eqnarray}
where $\hat{H}_j={\Omega(\mathbf{r}_j)}\hat{\sigma}^j_{21}/2+{\rm H.c.}$ is single atom Hamiltonian with $\hat{\sigma}^j_{\alpha\beta}=|\alpha^j\rangle\langle\beta^j|$ ($\alpha,\beta =1,\,2$). Here Rabi frequency $\Omega(\mathbf{r}_j)={d}_{21}\,\mathcal{E}(\mathbf{r}_j)$ depends on the slowly varying electric field $\mathcal{E}(\mathbf{r})$ and dipole moment ${d}_{21}$ between the Rydberg and groundstate. To be concrete, Cs atoms will be considered in this work as the respective dipole moment is relatively large compared to other alkali atoms (see Supplementary Material ({\bf SM})~\cite{SM} for details). Single-photon Rydberg excitation of ultracold Cs atoms has been demonstrated experimentally with nanosecond~\cite{tong_local_2004} and continuous lasers~\cite{hankin_two-atom_2014,jau_entangling_2016,wang_single-photon_2017,li_optical_2019}.
\begin{figure}
	\centering
	\includegraphics[width=0.48\textwidth]{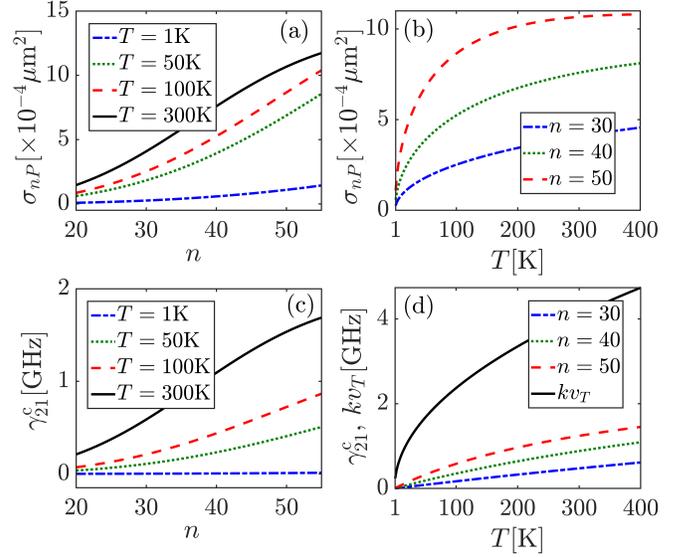}
	\caption{\footnotesize(Color online) \textbf{Collisional cross-section and decay rate in Rydberg $nP$ states}. The cross-section increases with higher $n$ (a) and temperature (b). Collisional decay rates monotonically increases with $n$ (c) and temperature (d). At room temperature, the rate is a few gigahertz for high Rydberg states that is comparable to the Doppler broadening ($kv_{{T}}$). Here the atomic density is ${\cal N}=5\times10^{15}{\rm cm}^{-3}$, and s-wave scattering length of Cs atoms $a_s\approx 21.7 a_B$  ($a_B$ the Bohr radius). }\label{collisonal}
\end{figure}

In addition to vdW and dipole-dipole interactions between Rydberg atoms, the attractive polarization interaction between electrons and groundstate atoms has been extensively studied previously~\cite{omont_theory_1977, beigman_collision_1995}. In ultracold gases, it leads to the formation of ultralong-range Rydberg molecules~\cite{greene_creation_2000,bendkowsky_observation_2009,li_homonuclear_2011,shaffer_ultracold_2018} and Rydberg polarons~\cite{camargo_creation_2018}. At high temperatures, it causes a spectra shift and inelastic collision due to mixing with other Rydberg states~\cite{beigman_collision_1995}. After compensating the shift with laser detuning, the inelastic collision is characterized by decay rate $\gamma_{21}^{\rm c}={\cal N} v_T\sigma_{nP}$~\cite{beigman_collision_1995} where  $v_T=\sqrt{2k_BT/M}$ is the thermal velocity ($M$ mass of Cs atoms), and $\sigma_{nP}$ the  collisional cross-section~\cite{SM}. As shown in Fig.~\ref{collisonal}(a) and (b), the cross-section becomes larger with increasing principal quantum number $n$ and temperature $T$. The decay rate moreover depends on atomic densities linearly. In high density ($>10^{15}\text{cm}^{-3}$) gases, the decay, e.g. $\gamma^{\text{c}}_{21}\sim 1$ gigahertz for $T=300$ K, is comparable to the Doppler broadening [Fig.~\ref{collisonal}(c)-(d)].

Taking into account the inelastic collision, dynamics of the system is described by a set of coupled Maxwell-Bloch equations~\cite{firstenberg_theory_2008}. In the following, we will focus on propagation of short pulses along $z$ direction while neglecting the diffraction as the medium is short. Applying the continuous density approximation, this yields the one dimensional (1D) Maxwell-Bloch equations,
\begin{subequations} \label{eq:bloch}
	\begin{eqnarray}
	&& i\frac{\partial }{\partial t}w(z)+\Omega(z)\rho_{12}(z)- \Omega^*(z)\rho_{21}(z)=0,\label{eq21} \\
	&& \left[i\frac{\partial }{\partial t}+i\gamma_{21}^{\rm c}-kv\right]
	\rho_{21}(z)+\frac{\Omega(z)}{2}w(z)-i\gamma_{21}^{\rm c}f(v)R_{21}(z) \nonumber\\
	&& -{\cal N}^{1/3}\int{dz^\prime dv^\prime f(v^\prime)V_r(z^\prime-z)\rho_{22,21}(z^\prime, z,t)}=0 \label{eq25}\\
	&&i\left(\frac{\partial}{\partial z}+\frac{1}{c}\frac{\partial}{\partial t} \right) \Omega(z)+\frac{k}{2}\chi(z,t)\Omega(z)=0,
	\end{eqnarray}
\end{subequations}
where $\rho_{\alpha\beta}(z)=\langle {\hat \sigma}_{\alpha\beta}(z) \rangle$ is the mean value of  operator ${\hat \sigma}_{\alpha\beta}(z)$, and $w(z)=1-2\rho_{22}(z)$ the population inversion. $R_{21}(z)=\int dv \rho_{21}(z,v,t) $ and $\chi(z)=2{\cal N} ( {d}_{12})^2\int dvf(v)\rho_{21}(z,v,t)/[\varepsilon_0\Omega(z)]$ are the integrated density and  susceptibility~\cite{firstenberg_theory_2008}, correspondingly. $f(v)=1/(\sqrt{\pi}v_T){\rm exp}[-(v/v_T)^2]$ is the 1D Maxwell-Boltzmann velocity distribution. These equations couple to two-body correlation $\rho_{\beta\alpha,\nu\mu}(z^\prime,z,t)\equiv\langle \hat{\sigma}_{\alpha\beta}
(z^\prime,t)\hat{\sigma}_{\mu\nu}(z,t)\rangle$, whose equation is cumbersome and given in \textbf{SM}~\cite{SM}. Note that spontaneous decay due to finite Rydberg lifetimes ($10\sim100 \mu$s) can be neglected in the dynamics due to mismatch of the time scales.

\begin{figure}
	\centering
	\includegraphics[width=0.47\textwidth]{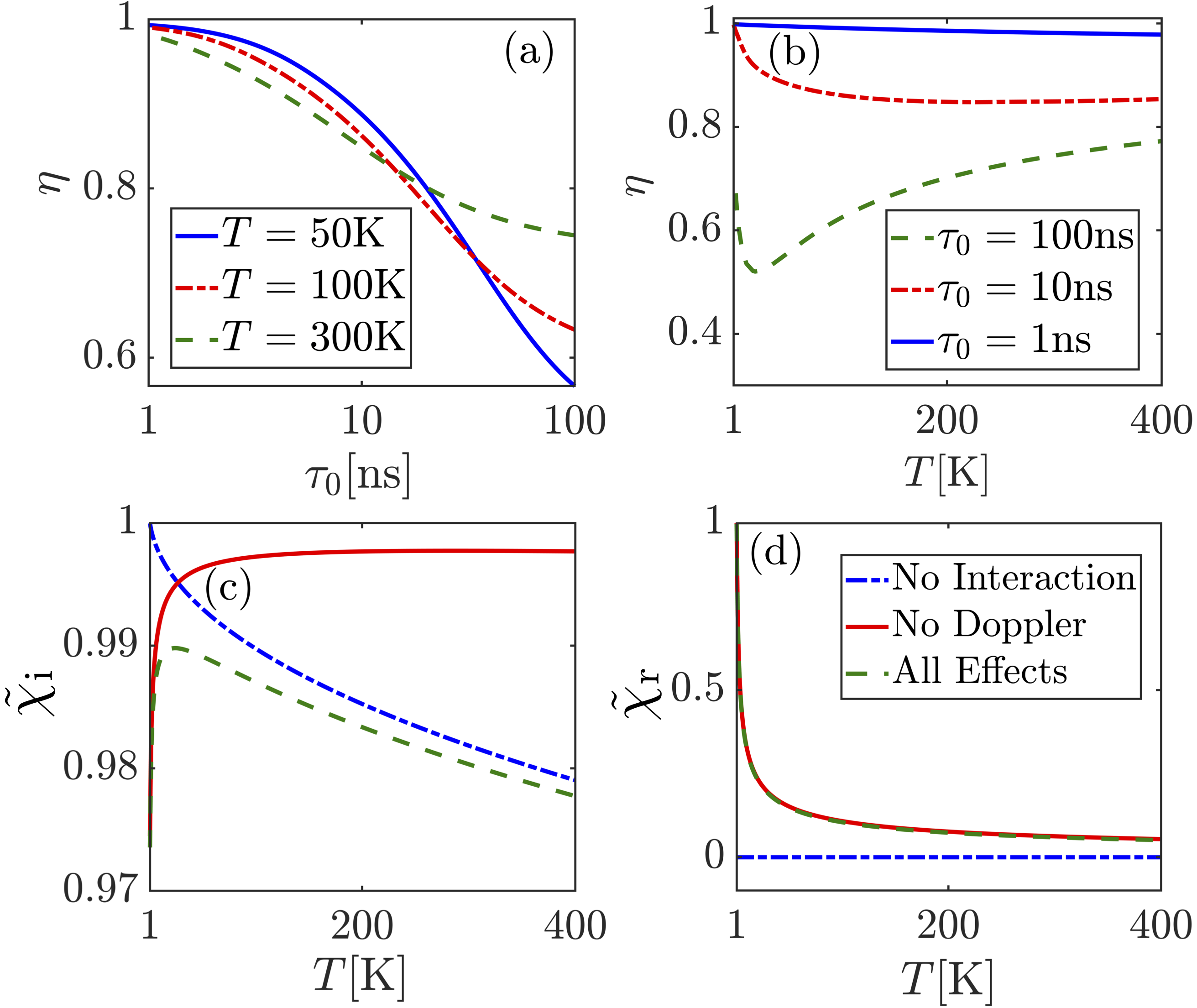}
	\caption{\footnotesize(Color online) \textbf{Transmission of the pulses}. The pulse duration (a) and temperature (b) of the medium affect the transmission. When $\tau\sim 1$ ns, $\eta\sim 1$ almost independent of the temperature. Notable absorption is found when $\tau\gg 1$ns in (a). Transmission varies with temperature non-monotonically for very long pulses, e. g. $\tau=100$ ns in panel (b). The imaginary  $\tilde{\chi}_{\text{i}}=\text{Im}[\tilde{\chi}(T)]/\text{Im}[\tilde{\chi}_s]$ (c) and real part $\tilde{\chi}_{\text{r}}=\text{Re}[\tilde{\chi}(T)]/\text{Re}[\tilde{\chi}_s]$ (d) of static susceptibility $\tilde{\chi}(T)$ at temperature $T$, scaled with respect to $\tilde{\chi}_s=\tilde{\chi}(T=1 {\rm K})$. The maximal $\tilde{\chi}_{\text{i}}$ and maximal absorption for $\tau=100$ ns in (b) both locate around $T=10$ K. The absorption is suppressed at low and high temperatures, due to less Rydberg excitations (hence decay) caused by the Rydberg blockade and Doppler effect~\cite{urvoy_optical_2013}, respectively. The Rydberg interaction gives large real part $\tilde{\chi}_{\text{r}}$, especially at low temperatures. The legend in (c) and (d) is same. We consider Rydberg state $|30P\rangle$ with lifetime $27.79\,\mu$s,  $L=400{\rm \mu m}$ and $\mathcal{N}=5\times10^{15}{\rm cm}^{-3}$.
	}\label{fig3}
\end{figure}
\textit{\textbf{Transmission of light pulses.}---}
We first study optical losses due to the collisional and Doppler effects. The former leads to dissipation directly while the  latter causes population partially trapped in Rydberg states, hence reducing the output intensity of the pulse after propagating in the medium (length $L$). To be concrete, we assume the pulse has a profile $\Omega(z=0)=\Omega_s\,{\rm sech}\left[(t-t_0)/\tau\right]$ with $\Omega_s$, $t_0$ and $\tau$ to be the amplitude, center and duration at the boundary $z=0$. We emphasize that the results in the following sections can be equally obtained by considering Gaussian pulses~\cite{SM}. The Maxwell-Bloch equation is solved by a combination of the 4th order Runge-Kutta and Chebyshev spectral method~\cite{boyd_chebyshev_2001}.

Using the spatial-temporal solution we evaluate transmission $\eta=\int^{+\infty}_{-\infty} dt |\Omega(L, t)|^2/\int^{+\infty}_{-\infty} dt |\Omega(0, t)|^2$ at the output $z=L$. For nanosecond pulses ($\tau\sim 1$ ns), we find that transmission $\eta\sim 1$, indicating that the medium is almost transparent  [Fig.~\ref{fig3}(a)]. An important feature is that transmission of nanosecond pulses is thermally robust. As shown in Fig.~\ref{fig3}(b), the reduction of $\eta$ is marginal when the temperature increases from $1$ K to $400$ K, though both the decay rate and Doppler broadening are a few gigahertz at $400$ K [see Fig.~\ref{collisonal}(c) and (d)].

For long pulses, transmission becomes smaller at higher temperatures [Fig.~\ref{fig3}(b)]. When $\tau\gg 10$ ns,  $\eta$ depends on the temperature non-trivially. For example, $\eta$ decreases and then increases with increasing temperature for $\tau=100$ ns, due to the interplay between the Doppler and collisional effect. We can understand this dependence qualitatively by examining static susceptibility $\tilde{\chi}(T)$ of infinitely long pulses, which is given analytically in \textbf{SM}~\cite{SM}. By analyzing the imaginary part of $\tilde{\chi}$ [Fig.~\ref{fig3}(c)], we find that the collisional decay (Doppler effect) plays leading roles at low (high) temperatures. Moreover the real part of $\tilde{\chi}$ is large at lower temperatures [Fig.~\ref{fig3}(d)]. This means that the pulse can gain an optical phase during propagation.

\textit{\textbf{Rydberg-SIT of nanosecond pulses.}---} In the following, we will focus on the high transmission situations, where so-called \textit{self-induced transparency}~\cite{mccall_self-induced_1967,lamb_analytical_1971} can form.
Without atom-atom interactions, SIT occurs if areas of the input pulse  $\theta(z)=\int_{-\infty}^{\infty}\Omega(z,t)dt=\Omega_s\tau\pi$ are multiple of $2\pi$, i.e. $\Omega_s\tau$ is an even number, governed by the area theorem~\cite{lamb_analytical_1971}. This nonlinear effect is rooted solely from high light intensities, which reshape the pulse into a stable, bright soliton, i.e. no absorption or distortion. The nonlinearity reduces the group velocity [$v_g\approx 2\varepsilon_0|\Omega|^2/(k\mathcal{N}{d}_{12}^2)$] but does not affect optical phases~\cite{lamb_analytical_1971}.

\begin{figure}
	\centering
	\includegraphics[width=0.49\textwidth]{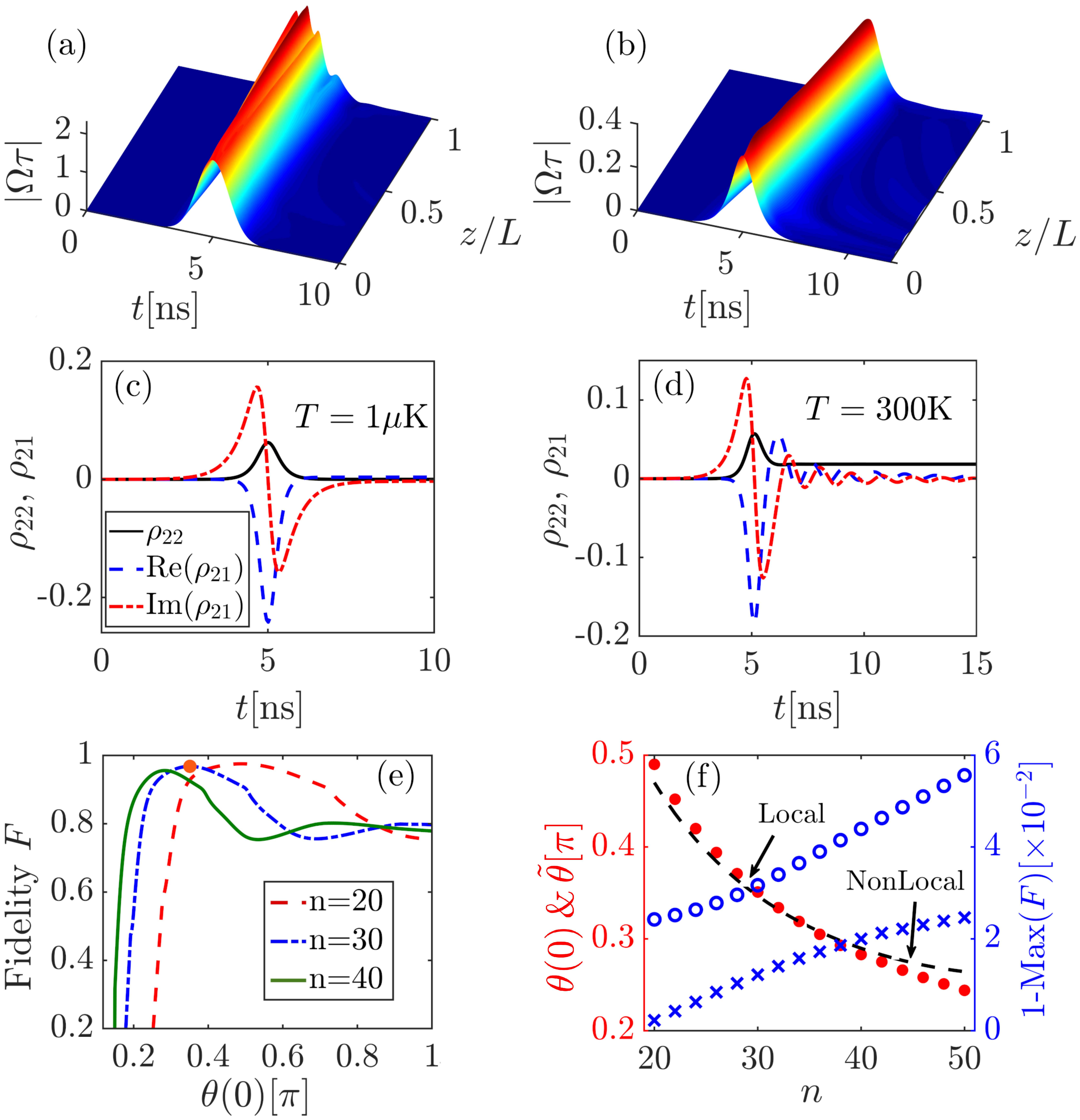}\\
	\caption{\footnotesize(Color online) \textbf{Rydberg-SIT and the optimal area}. The pulse is distorted when $\theta(0)=2\pi$ (a) and stable when $\theta(0)=0.35\pi$ (b), corresponding to Rydberg-SIT. The transient dynamics of atoms at $T=1\,\mu$K (c) and $300$ K (d).  $\rm{Im}(\rho_{21})$ is symmetric with respect to $t_0$. Note that the time average of $\rm{Im}(\rho_{21})$ is negligibly small at $300$ K, which is important to the formation of Rydberg-SIT. (e) Fidelity $F$ as a function of initial area $\theta(0)$. Maximal fidelities appear at Rydberg-state dependent optimal areas. (f) Optimal area (filled circle) and corresponding fidelity at $1\, \mu$K (star) and $300$ K (empty circle). The MF (dashed) and numerical calculation agree. In panel (a)-(d) $n=30$. In (e)-(f), the pulse area is varied by changing $\Omega_s$.  Other parameters $t_0=5$~ns, $\tau=1$~ns, $L=400~{\rm \mu m}$ and $T=300$~K.
	}\label{fidelity}
\end{figure}

Due to the strong Rydberg interaction, the pulse profile is distorted when the input area $\theta(0)=2\pi$ [Fig.~\ref{fidelity}(a)], while is preserved if $\theta(0)=0.35\pi$ [Fig.~\ref{fidelity}(b)], giving rise to \textit{Rydberg-SIT}. Similar to SIT, the formation of Rydberg-SIT can be understood by analyzing the atomic dynamics~\cite{mccall_self-induced_1967,lamb_analytical_1971}. The dynamics is independent of $z$ since the nanosecond pulse translates in the medium. Crucially important for Rydberg-SIT is that coherence $\rm{Im}(\rho_{21})$ is symmetric with respect to $t_0$, i.e. positive (negative) when $t<t_0$ ($t>t_0$) [Fig.~\ref{fidelity}(c)]. As $\rho_{22},\,\rho_{21}\to 0$ when $t\to +\infty$, the light is thus absorbed and then emitted coherently. When $T$ increases from $1\,\mu$K [Fig.~\ref{fidelity}(c)] to $300$ K [Fig.~\ref{fidelity}(d)], modifications of the dynamics are marginal. Such transient dynamics guarantees the formation of Rydberg-SIT at the optimal area $\theta(0)=0.35\pi$.

We define fidelity $F= |\int^{+\infty}_{-\infty}dt\,\Omega(L, t)\Omega(0, t)|^2/\int_{-\infty}^{+\infty}dt\,|\Omega(L, t)|^2\int_{-\infty}^{+\infty}dt\,|\Omega(0, t)|^2$ to quantify the deformation. $F=1$ if the input and output pulse are identical. When $0<\theta(0)\le 2\pi$, $F$ indeed displays a single maximal at $\theta(0)=0.35\pi$. For different Rydberg states, optimal areas vary with $n$, see Fig.~\ref{fidelity}(e).
We furthermore carry out large scale calculations for $20\le n\le 50$. It is found that the optimal area  decreases monotonically with increasing $n$, while the corresponding deformation fidelity is high [Fig.~\ref{fidelity}(f)]. Note that Rydberg-SIT can also be achieved with Gaussian pulses, which lead to similar optimal areas and fidelities as shown in \textbf{SM}~\cite{SM}.

\textit{\textbf{State-dependent optimal areas.}---} \label{analytical}
Inspired by the transient dynamics of Rydberg-SIT [Fig.~\ref{fidelity}(c)-(d)] we will develop a mean field (MF) theory for the Bloch equation (BE) to understand the optimal area. To deal with the two-body interaction term in Eq.~(\ref{eq25}), we apply a local field approximation to the two-body correlations, i.e. $\rho_{22,21}(z',z,t)\approx \rho_{22}(z,t)\rho_{21}(z,t)$~\cite{sevincli_nonlocal_2011} as the pulse is much longer than ranges of the Rydberg interaction. With this approximation Eq.~(\ref{eq25}) becomes,
\begin{equation}
\frac{\partial\rho_{21}}{\partial t}\approx -\gamma_{21}^{\rm c}[\rho_{21}-f(v)R_{21}]-i(kv+u{\rho}_{22})
{\rho}_{21}-\frac{i{\Omega}w}{2},\label{mfblochc}
\end{equation}
while other equations are not affected formally. Here $u= 2\mathcal{N}^{1/3}\int^{\infty}_{0}dz V(z)$ is an effective  Rydberg interaction. To avoid divergence in the integral, the vdW potential is modified at short distances to have a soft-core shape, $V(z)\approx C_6/(z^6+z_m^6)$, when atomic distances are smaller than the blockade radius $z_m=(|C_6|/\Omega_s)^{1/6}$~\cite{lukin_dipole_2001}. This allows us to analytically evaluate the effective interaction $u=4\pi {\cal N}^{1/3}C_6^{1/6}\Omega_s^{5/6}/3$, which depends on the density, Rabi frequency and Rydberg state.

Depending on the ratio $(kv_T+ u)/\gamma_{21}^{\rm{c}}$, three different regimes of the coherence are obtained approximately according to Eq.~(\ref{mfblochc}). Fixing $T$, a Doppler broadening dominant region appears at low densities when $kv_T> u\gg\gamma_{21}^{\rm{c}}$, as shown in Fig.~\ref{fig1}(c).  For sufficiently high densities [dotted line in Fig.~\ref{fig1}(c) with $10kv_T=u$] Rydberg interactions overtake the other two effects, i.e. $u>kv_T\gg \gamma_{21}^{\rm{c}}$. This is the most interesting region where Rydberg-SIT can form. Further increasing densities (dashed line, $kv_T+u=100\gamma_{21}^{\rm{c}}$), the collisional decay starts to kick in and causes losses. The overall decay will also depend on the propagation distance.

In the next, we will find the optimal areas analytically in the Rydberg interaction dominant region (by neglecting terms involving $kv_{T}$ and $\gamma_{21}^{\text{c}}$). As the nonlinear Eq.~(\ref{mfblochc}) is difficult to integrate even with this approximation, we will apply the following ansatz solution ${\rho}_{22} = A[1-\cos\int_{-\infty}^t\Omega_0(t')dt']$ and ${\rho}_{21} = -\frac{iB}{2}\cos\int_{-\infty}^t\Omega_0(t') dt' + C\rho_{22}$ where $A$, $B$, and $C$ are trial parameters, and $\Omega_0(t)=\sqrt{2\pi}\exp\left(-t^2/2\tau^2\right)/\tau$. Such ansatz ensures the symmetry of the transient dynamics, i.e. $\text{Im}[\rho_{21}]$ is symmetric with respect to the pulse center, and $\rho_{12}=\rho_{22}=0$ when $t\to \infty$. We then approximate the pulse $\Omega$ in the MF equation with a Gaussian ${\Omega} = \tilde{\theta}\exp\left(-t^2/2\tau^2\right)/\sqrt{2\pi}\tau$ where  $\tilde{\theta}$ is the optimal area to be determined.

Substituting the ansatz to the MF equation, the trial parameters and area $\tilde{\theta}$ can be calculated analytically (see \textbf{SM}~\cite{SM} for details). Explicitly, the Rydberg-state-dependent area is given by
\begin{equation}\label{mfarea}
\tilde{\theta}=\frac{2\pi}{u\tau}\left(\sqrt{2\pi u^2\tau^2+\pi^2}-\pi\right)^{\frac{1}{2}},
\end{equation}
which is the key result of the MF calculation. Eq.~(\ref{mfarea}) shows $\tilde{\theta}\to 2\pi$ when $u\to 0$, recovering the area theorem in non-interacting SIT~\cite{lamb_analytical_1971}. Increasing $u$, $\tilde{\theta}$ decreases gradually. When compared with numerical data, an excellent agreement is found  if $n<40$. Small deviations for $n>40$ attribute to the local field approximation, which can not fully capture the long-range vdW interaction in highly excited Rydberg states.

\textit{\textbf{Conclusion and discussion.}---}
In this work, we have studied the propagation dynamics of nanosecond pulses in thermal, high-density Rydberg gases.  We have shown that strong dispersive optical nonlinearities can be achieved from low to high temperatures. Due to the Rydberg nonlinearity, Rydberg-SIT can form in thermal atomic gases which is largely immune to the Doppler broadening and collisional decay. A key finding is that the optimal area of  Rydberg-SIT is reduced by the Rydberg atom interaction. The optimal area and its dependence on the interaction are determined both numerically and analytically.

This work opens exciting opportunities to study nonlinear optics and to implement quantum information processing at nanosecond time scales with warm Rydberg gases. Beyond the current level scheme, one could exploit strong Rydberg nonlinearities of short laser pulses using, e.g. two-photon excitations. This allows to investigate the formation and stability of simultons~\cite{Ogden_Quasisimultons_2019} in the presence of strong Rydberg atom interactions. The dispersive Rydberg nonlinearity could be used in quantum information processing, such as to build optical phase gates~\cite{agrawal_nonlinear_2012,islam_ultrafast_1989, mcleod_1d_1995,kochetov_logic_2019} with Rydberg-SIT.

\begin{acknowledgements}
	We thank insightful discussions with Thomas Gallagher, Igor Lesanovsky, Sebastian Slama, Lin Li, Jing Zhang, Jianming Zhao, Junmin Wang and Stephen Hogan.  ZB and GH acknowledge National Science Foundation (NSF) (11904104, 11975098, 11174080, 11847221), The Shanghai Sailing Program (18YF1407100), China Postdoctoral Science Foundation (2017M620140), and the International Postdoctoral Exchange Fellowship Program (20180040). CSA acknowledges financial support from EPSRC Grant Ref. Nos. EP/R002061/1, EP/M014398/1, EP/S015973/1, EP/R035482/1 and EP/P012000/1, the EU RIA
	project `RYSQ' project, EU-H2020-FETPROACT-2014 184 Grant No. 640378
	(RYSQ) and DSTL. W. L. acknowledges support from the EPSRC through grant No. EP/R04340X/1 via the QuantERA project “ERyQSenS”, the Royal Society grant No. IEC$\backslash$NSFC$\backslash$181078, and the UKIERI-UGC Thematic Partnership (IND/CONT/G/16-17/73).
\end{acknowledgements}

	\bibliography{References}
	
	\clearpage
	\begin{widetext}
		
		\begin{center}
			{\Large Supplementary information for}
		\end{center}
		\begin{center}
			{\Large \bf Self-induced Transparency in Warm and Strongly Interacting Rydberg Gases}
		\end{center}
		\begin{center}
			{\large Zhengyang Bai, Charles S. Adams, Guoxiang Huang and Weibin Li}
		\end{center}
		\vspace{10mm}
		\setcounter{secnumdepth}{2}
		\setcounter{equation}{0}
		\setcounter{figure}{0}
		\setcounter{table}{0}
		\setcounter{page}{1}
		\makeatletter
		
		\renewcommand{\theequation}{S\arabic{equation}}
		\renewcommand{\thefigure}{S\arabic{figure}}
		\renewcommand{\bibnumfmt}[1]{[S#1]}
		%
		
		
		This supplementary material gives further details on the analysis in the main text.
		
\section*{A. The properties of Rydberg atoms}
%
We shall first calculate the inelastic collision between Rydberg atom and groundstate atom. In dense and high temperature gases, the electron-atom collision mixes energetically neighboring Rydberg states. Effectively, the resulting \textit{inelastic collision} causes decay of the Rydberg state with rate $\gamma_{21}^{\rm c}={\cal N} v_T\sigma_{nP}$~\cite{beigman_collision_1995} with $\cal{N}$ and $v_T=\sqrt{2k_BT/M}$ to be the density  and thermal velocity ($M$ mass of Cs atoms). The inelastic cross-section $\sigma_{nP}$ is~\cite{lebedev_nonresonant_1985,beigman_collision_1995}
\begin{equation}
\sigma_{nP}=\sum_{n'}\frac{4v_0^2a_s^2}{v_T^2n'^3}\left[\arctan\left(\frac{2}{\lambda}\right)-\frac{\lambda}{2}\ln\left(\frac{4+\lambda^2}{\lambda^2}\right)\right], \nonumber
\end{equation}
where $\lambda = |\delta_P + n'-n|v_0/(n^2v_T)$ characterizes the inelasticity for the  $nP\to n'$ manifold transition.  $\delta_P$ and $v_0=\hbar/(m a_b)$  are the quantum defect and orbital velocity of the Rydberg electron (mass $m$). The non-trivial dependence of the cross-section on temperature and Rydberg states is shown in Fig.~2a and 2b in the main text. Note that cross-sections in Rydberg states are smaller than that of the low-lying states (e.g. $|6P\rangle$)~\cite{steck_alkali_nodate}, which could cause stronger decay.

The strength of the single photon transition depends on the transition dipole moment ${d}_{21}$ between the ground state and Rydberg state $|nP_{3/2}\rangle$.  Fig.~\ref{Dipole} show results for Rb and Cs atoms. The transition dipole for Cs is relatively large. We shall note that, to produce a strong Rydberg nonlinearity, the input light intensity is needed to be intense due to the fact that the transition dipole moment ${d}_{21}$ between the ground state and Rydberg state is small, but it is still available with current experimental capabilities~\cite{tong_local_2004}. To be concrete, we can estimate the peak intensity $I_{\rm max}=c\epsilon_0|E_{p {\rm max}}|^2/2\simeq0.08$ MW cm$^{-2}$ for generating such Rydberg-SIT in Fig.~4d in the main text.

\section*{B. Dynamics Of The Rydberg Atom}
In this section, we provide the theory of the many-body dynamics in  warm and strongly interacting Rydberg gases. We will focus on the derivation of the hierarchy of equations for many-body correlators provided in
the main text. We first treat the dynamics under the effective Hamiltonian in the Heisenberg picture. To this end we obtain the equation of motion for the expectation values $\hat{\rho}$ from the corresponding operator
Heisenberg equations. This yields the explicit expression of the Bloch equation,
\begin{figure}
	\centering
	\includegraphics[width=0.45\textwidth]{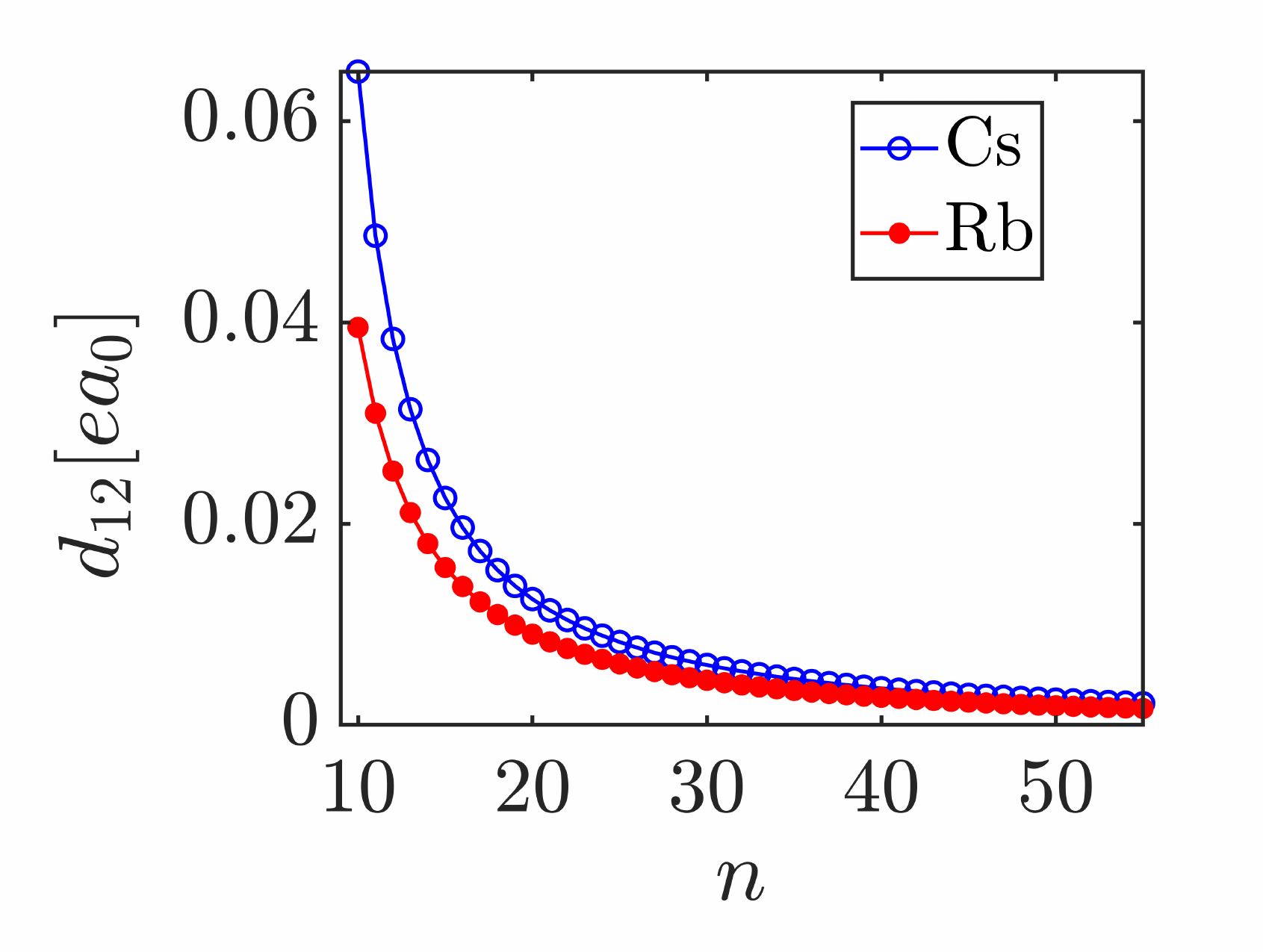}\\
	\caption{\footnotesize(Color online) The transition dipole moment ${d}_{21}$ between the ground state and Rydberg state $|nP_{3/2}\rangle$ for Rb and Cs atom.
	}\label{Dipole}
\end{figure}
\begin{subequations} \label{eq:bloch}
	\begin{eqnarray}
	&& i\frac{\partial }{\partial t}\rho_{11}(z)-i\Gamma\rho_{22}(z)-\text{Re}\left[\Omega(z)\rho_{12}(z)\right]=0,\label{eq21} \\
	&& \frac{\partial }{\partial t}\rho_{22}(z)+\Gamma\rho_{22}(z)+\text{Im}\left[\Omega(z)
	\rho_{12}(z)\right]=0,\label{eq22}\\
	&& \left[i\frac{\partial }{\partial t}+i\gamma_{21}^{\rm c}-kv\right]
	\rho_{21}(z)+\frac{\Omega(z)}{2}w(z)-i\gamma_{21}^{\rm c}f(v)R_{21}(z) \nonumber\\
	&& -{\cal N}\int{d^3r^\prime dv^\prime f(v^\prime)V(r^\prime-r)\rho_{22,21}(r^\prime, r,t)}=0,\label{eq23}
	\end{eqnarray}
\end{subequations}
where $\rho_{\alpha\beta}(z)=\langle {\hat \sigma}_{\alpha\beta}(z) \rangle$ is the mean value of  operator ${\hat \sigma}_{\alpha\beta}(z)$, and these equations couple to two-body correlation $\rho_{\beta\alpha,\nu\mu}(z^\prime,z,t)\equiv\langle \hat{\sigma}_{\alpha\beta}
(z^\prime,t)\hat{\sigma}_{\mu\nu}(z,t)\rangle$, which are contributed from the interaction between Rydberg atoms. $\Gamma$ is the spontaneous decay rate in the Rydberg state, which is a small quantity, i.e. $\Gamma\ll \Omega, \gamma_{21}^{\text{c}}, kv_T$.

From equation (\ref{eq23}), we see that to solve the one-body  correlators $\rho_{\beta\alpha}(\mathbf{r},v,t)\equiv\langle {\hat \sigma}_{\alpha\beta}(\mathbf{r},t)\rangle$, one needs to solve the equations of motion for the two-body correlators $\rho_{\beta\alpha,\nu\mu}(\mathbf{r}^\prime,\mathbf{r},$ $v^\prime,v,t)\equiv\langle \hat{\sigma}_{\alpha\beta}
(\mathbf{r}^\prime,v^\prime,t)\hat{\sigma}_{\mu\nu}(\mathbf{r},v,t)\rangle$, where a complete list of the equations of motion for the two-body density matrix (correlator) elements of the system is given by
\begin{subequations}
	\begin{eqnarray}\label{operator3}
	&&\left(i\frac{\partial }{\partial t}+d_{21}(v^\prime)+d_{21}(v)- V_r(\mathbf{r}^\prime-\mathbf{r})\right)\rho_{21,21}-2\Omega\rho_{22,21}+\Omega\rho_{21}=0, \\
	&&\left(i\frac{\partial }{\partial t}+d_{21}(v)+d_{12}(v^\prime)\right)\rho_{21,12}-\Omega\rho_{22,12}+\Omega^\ast\rho_{21,22}+\frac{\Omega}{2}\rho_{12}-\frac{\Omega^\ast}{2}\rho_{21}=0, \\
	&&\left(i\frac{\partial }{\partial t}+i\Gamma+d_{21}(v)- V_r(\mathbf{r}^\prime-\mathbf{r})\right)\rho_{22,21}+\frac{\Omega}{2}\rho_{12,21}-\frac{\Omega^\ast}{2}\rho_{21,21}-\Omega\rho_{22,22}+\frac{\Omega}{2}\rho_{22}=0,\nonumber\\
	&&\\
	&&\left(i\frac{\partial }{\partial t}+2i\Gamma\right)\rho_{22,22}+\Omega\rho_{12,22}-\Omega^\ast\rho_{21,22}=0,
	\end{eqnarray}
\end{subequations}
where $d_{21}(v)=i\gamma_{21}^c-kv$ and $d_{12}(v)=i\gamma_{21}^c+kv$. Here the motion of equations are closed for one-body and two-body correlations and their number becomes finite, due to the fact that higher-order many-body correlators involved in the hierarchy play no significant role and hence can be truncated in our calculation~\cite{bai_stable_2019}.
\begin{figure}
	\centering
	\includegraphics[width=0.8\textwidth]{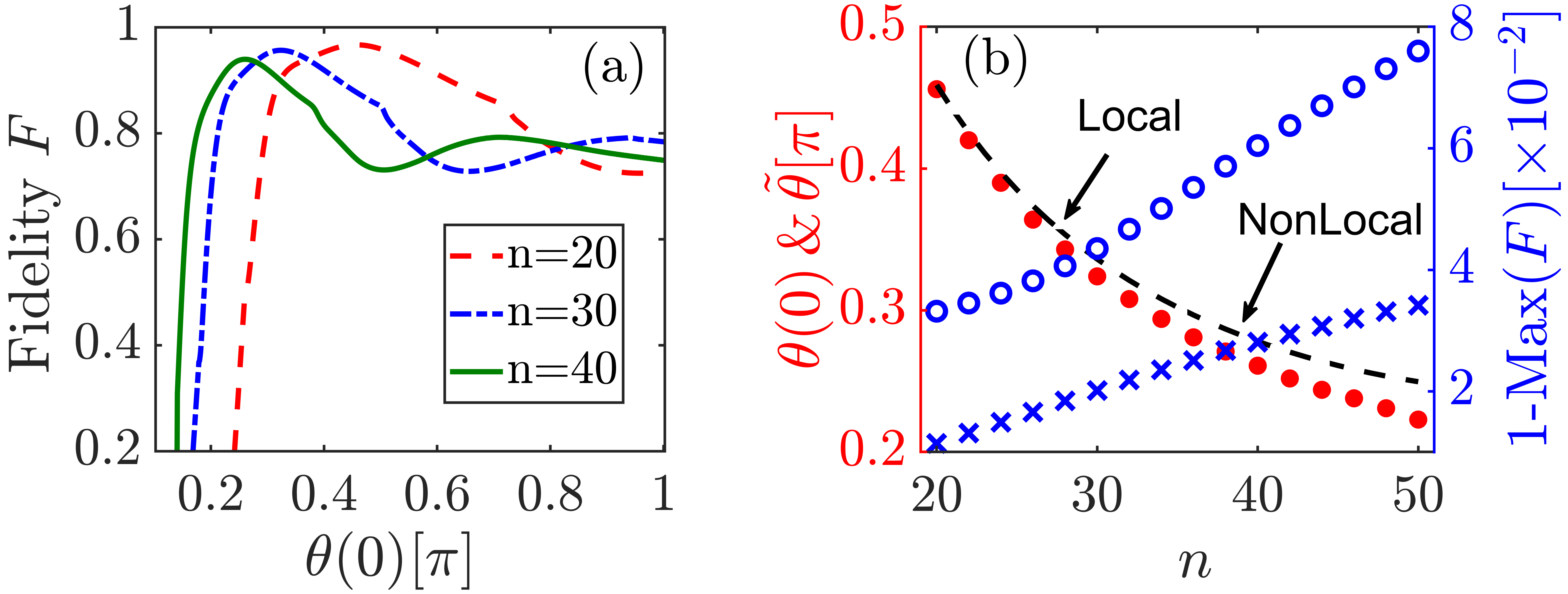}\\
	\caption{\footnotesize(Color online) The optimal area for light pulse with Gaussian profile. (a) Fidelity $F$ as a function of initial area $\theta(0)$. Maximal fidelities appear at areas depending on the Rydberg states. The pulse area is varied by changing the amplitude of $\Omega$. (b) The optimal area (circle) and corresponding output fidelity (square) of different Rydberg states. The dashed line is the optimal area predicted by Eq.~(3) in the main text, which agrees with our numerical data. The parameters used here are same as the one in Fig.~4 in the main text.
	}\label{Gaussian}
\end{figure}

Focusing on light propagation in $z$-axis (i.e. neglecting diffraction), the propagation
of the light pulse field is governed by one dimensional (1D) Maxwell equation,
\begin{eqnarray}\label{Max2}
&&i\left(\frac{\partial}{\partial z}+\frac{1}{c}\frac{\partial}{\partial t} \right) \Omega(z)+\frac{k}{2}\chi(z,t)\Omega(z)=0,
\end{eqnarray}
where $\chi(z)=2{\cal N} ( {d}_{12})^2\int dvf(v)\rho_{21}(z,v,t)/[\varepsilon_0\Omega(z)]$ are the integrated density and temperature dependent susceptibility~\cite{firstenberg_theory_2008}, correspondingly. $f(v)=1/(\sqrt{\pi}v_T){\rm exp}[-(v/v_T)^2]$ is the 1D Maxwell-Boltzmann velocity distribution. In the main text, we use a combination of the 4th order Runge-Kutta and Chebyshev spectral method to solve the coupled Maxwell-Bloch equation~\cite{boyd_chebyshev_2001}.

In the main text, we have already shown numerical results when the profile is described by the sech function. Without loss of generality, we show the results for the Gaussian profile in Fig.~\ref{Gaussian}. Similar optimal areas can be found in the numerical calculations. This also justifies that it is a good approximation to use the Gaussian profile in the mean field calculation.

\section*{C. Steady-state susceptibility}
In this section, we shall explain absorption for long pulses. We assume that a gate excitation is stored initially~\cite{gorshkov_photon-photon_2011}. After preparing the initial state, a source field is sent to the medium and collide with the gate Rydberg excitation. The interaction between gate and source gatoms reads $V_d(z_{jk})=-C_6^d/|z_j-z_k|^6$ ($C_6^d$ dispersion coefficient of the interstate interaction). To obtain an analytical expression of coherence $\rho_{21}$, we assume that the light is a continuous wave, thus optical bloch equation can be easily written in the frequency space,
\begin{subequations} \label{twoatom}
	\begin{eqnarray}
	&& -i\omega\tilde{\rho}_{22}(z)+\Gamma\tilde{\rho}_{22}(z)+\text{Im}\left[\Omega(z)
	\tilde{\rho}_{12}(z)\right]=0,\label{eq22}\\
	&& \left(\omega+i\gamma_{21}^c-V_d(z)-kv\right)
	\tilde{\rho}_{21}(z)-\Omega(z)\tilde{\rho}_{22}(z)+\Omega(z)/2=0,
	\end{eqnarray}
\end{subequations}
where $\rho_{\alpha\beta}=\int d\omega\tilde{\rho}_{\alpha\beta}{\rm exp}(-i\omega t)$.
To the lowest order of the frequency, we derive the coherence $\tilde{\rho}_{21}$
\begin{eqnarray}\label{rho21}
\tilde{\rho}_{21}=\frac{(i\gamma_{21}^c+V_d+kv)\Gamma\Omega}{2[\gamma_{21}^c(\Omega^2+\gamma_{21}^c\Gamma)+\Gamma(V_d+kv)^2]},
\end{eqnarray}
After substituting Eq.~(\ref{rho21}) into the Maxwell Equation (\ref{Max2}), and then integrate it, the coupled Bloch-Maxwell equation can be solved,
\begin{subequations}\label{chi}
	\begin{eqnarray}
	&& \Omega(z)=\Omega(0){\rm exp}(i\phi-\alpha),\label{chia}\\
	&& \tilde{\chi}=\frac{{\cal N}{d}_{12}^2}{2\varepsilon_0}\int f(v)\frac{(i\gamma_{21}^c+V_d+kv)\Gamma}{\gamma_{21}^c(\Omega^2+{\gamma_{21}^c}\Gamma)+\Gamma(V_d+kv)^2}dv,\label{chib}
	\end{eqnarray}
\end{subequations}
where $\phi=k\int_0^L{\rm Re}(\tilde{\chi})dz$ characterizes the accumulated phase and $\alpha=k\int_0^L{\rm Im}(\tilde{\chi})dz$ contributes to the optical absorption. Fig.~2c in the main text shows the imaginary of $\tilde{\chi}$, and one can see that the absorption is suppressed at low and high temperatures, due to less Rydberg excitations (hence decay) caused by the Rydberg blockade and Doppler effect, respectively.

For nanosecond pulses ($\tau \sim1$ ns), we find that transmission $\eta\sim 1$, indicating that the medium is almost transparency (see Fig.~2a in the main text). Moreover, the real part of $\tilde{\chi}$ is nonzero (see Fig. 3d in the main text).

\section*{D. Mean field calculation}
In the MF approximation, the Doppler Broadening and collisional decay terms are neglected. For Rydberg-SIT, the pulse is translationally invariant in the medium, we can neglect the spatial dependence in the BE. This gives the following MF equation,
\begin{eqnarray}
\dot{w} &=& i\Omega(\rho_{21}-\rho_{12}),\label{eq:pop}\\
\dot{\rho}_{12} &\approx& \frac{i}{2}\Omega w + \frac{iu}{2}(1-w)\rho_{12}, \\
\dot{\rho}_{21} &\approx &-i \frac{i}{2}\Omega w -\frac{iu}{2}(1-w)\rho_{21},
\end{eqnarray}
where the pulse is given by ${\Omega} = \tilde{\theta}\exp\left(-t^2/2\tau^2\right)/\sqrt{2\pi}\tau$. We can easily show that the area of the pulse is $\tilde{\theta}$. To be concrete, we have assumed that the MF equation is dealt at $z=0$, such that the phase of the pulse is negligible, i.e. $\Omega$ is a real function.

For convenience, we define two new variables for the coherence, $s = \rho_{12}+\rho_{21}$ and $h = \rho_{12}-\rho_{21}$. The MF equations can be rewritten as
\begin{eqnarray}
\dot{w} &=& i\Omega h, \label{eq:w} \\
\dot{s} &=& \frac{iu}{2}(1-w)h, \label{eq:cohreal}\\
\dot{h} &=& i\Omega w +\frac{iu}{2}(1-w)s. \label{eq:cohimag}
\end{eqnarray}

We will solve the MF equations with the ansatz
\begin{eqnarray}
{\rho}_{22} &&= A[1-\cos\int_{-\infty}^t\Omega_0(t')dt'],\label{eq:ansatz1}\\
{\rho}_{21} &&= -\frac{iB}{2}\cos\int_{-\infty}^t\Omega_0(t') dt' + C\rho_{22},\label{eq:ansatz2}
\end{eqnarray}
where $A$, $B$, and $C$ are trial parameters. In the ansatz, we choose a Gaussian pulse $\Omega_0(t)=\theta_0\exp\left(-t^2/2\tau^2\right)/\tau\sqrt{2\pi}$, whose area is $\theta_0=2\pi$.

To solve the MF equation, we first substituting the ansatz into Eq.~(\ref{eq:w}) and find
\begin{eqnarray}
\dot{w} =-2A\Omega_0\sin F(t) = -B\Omega\sin F(t),
\end{eqnarray}
where we have defined,
\begin{equation}
F(t)=\int_{-\infty}^t \Omega_0 dt'=\frac{\theta_0}{2}\left[1+\text{Erf}\left(\frac{t}{\sqrt{2}\tau}\right)\right].
\end{equation}
One can prove that $\cos F(t)=-1$ when $t=0$ and $\cos F(t)=1$ when $t=\pm \infty$. If we cancel the time-dependent terms on both sides in the above equation, this gives a relation between $A$ and $B$,
\begin{eqnarray}
A=\frac{\tilde{\theta}}{2\theta_0}B. \label{eq:A}
\end{eqnarray}

From Eq.~(\ref{eq:cohreal}) and $\dot{s}=4AC\Omega_0\sin F(t)$, we can get a time-dependent equation
\begin{equation}
4AC\Omega_0 =-uAB[1-\cos F(t)].
\end{equation}
We consider $t=0$ in this equation. Using $\cos F(0)=-1$ and $A\neq 0$, we obtain a relation between $B$ and $C$,
\begin{equation}
C=\frac{\sqrt{2\pi}u\tau}{2\theta_0}B. \label{eq:C}
\end{equation}

Carrying out same procedure for Eq.~(\ref{eq:cohimag})  and using $\dot{h} = iB\Omega_0\cos F(t)$, we obtain an equation
\begin{equation}
B\Omega_0\cos F(t) = \Omega[1-2A(1-\cos F(t))] + 4uA^2C(1-\cos F(t))^2. \label{eq:mfimag}
\end{equation}
At $t=+\infty$, we can apply $\cos F(t)=1$ and $\lim_{t\to\infty}\Omega/\Omega_0=\tilde{\theta}/\theta_0$. This allows us to derive
\begin{equation}
B=\frac{\tilde{\theta}}{\theta_0}. \label{eq:B}
\end{equation}

To solve the parameters, we again use Eq.~(\ref{eq:mfimag}) and consider $t=0$. This leads to equation
\begin{equation}
B\theta_0+\tilde{\theta}(1-4A)+16u\tau CA^2=0.
\end{equation}
Substituting Eqns.~(\ref{eq:A}), (\ref{eq:B}) and (\ref{eq:C}) into this equation, this gives an equation for  $\tilde{\theta}$
\begin{equation}
\frac{u^2\tau^2}{\theta_0^5} \tilde{\theta}^4 +\frac{1}{\theta_0^2}\tilde{\theta}^2 -1=0.
\end{equation}
Solving this equation, we get the solution to the area,
\begin{equation}
\tilde{\theta}=\frac{2\pi}{u\tau}\left(\sqrt{2\pi u^2\tau^2+\pi^2}-\pi\right)^{\frac{1}{2}},
\end{equation}
which was used to compare against numerical data in the main text. With the explicit solution, parameters $B,\,A$ and $C$ can also be solved one by one.
	\end{widetext}
	
\end{document}